%% file: main.tex
\documentclass[a4paper]{llncs}

\usepackage{graphicx}
\usepackage{hyperref}
\usepackage{listings}
\usepackage{setspace}
\usepackage{times}
\usepackage{url}
\usepackage{wrapfig}
\usepackage{xcolor}
\usepackage{xspace}

\usepackage{tikz}
\usetikzlibrary{positioning, automata, arrows, calc, shapes, patterns}

\renewcommand\subparagraph{\typeout{LLNCS warning: You should not use
    \string\subparagraph\space with this class}\vskip0.5cm
  You should not use \verb|\subparagraph| with this class.\vskip0.5cm}

\usepackage{titlesec}

\hypersetup{
     bookmarks    = true,
     colorlinks   = true,
     citecolor    = blue,
     urlcolor     = blue,
     linkcolor    = blue,
     backref      = true,
     urlbordercolor = white,
     pdfborder    = 0 0 0,
     hyperindex   = true,
     hyperfootnotes = false,
}

\newcommand{\ignore}[1]{{}}
\newcommand{\nusmv}{\textsc{NuSMV}\xspace}
\newcommand{\nuxmv}{\textsc{nuXmv}\xspace}
\newcommand{\xsap}{\textsc{xSAP}\xspace}
\newcommand{\fsap}{\textsc{FSAP}\xspace}

\newcommand{\smt}{\textsc{SMT}\xspace}

\newcommand{\myurl}[1]{{\small\url{#1}}}

\lstnewenvironment{codesnippet}[2]{\lstset{language=C++,
    morekeywords={MODULE,ASSIGN,TRANS,INIT,INVAR,IVAR,INVARSPEC,
                  init,main,next,case,esac,VAR,Real,real,Integer,integer,
                  word,Word,array,TRUE,FALSE},
    backgroundcolor=\color{gray!10},
    keywordstyle=\color{blue!55}\bfseries,
    basicstyle=\tiny\sffamily,
    commentstyle=\color{green!60!black},
    stringstyle=\rmfamily,
    numbers=left,
    numberstyle=\tiny\ttfamily,
    numbersep=0.8em,
    showstringspaces=false,
    breaklines=false,
    frame=lines,
    float=!t,
    captionpos=b,
    framexleftmargin=2.5em,
    xleftmargin=2.5em,
    caption={#1},
    label={#2}
}}{}

\title{The \xsap Safety Analysis Platform}
\author{
B.~Bittner \and
M.~Bozzano \and
R.~Cavada \and
A.~Cimatti \and\\
M.~Gario \and
A.~Griggio \and
C.~Mattarei \and
A.~Micheli \and
G.~Zampedri}
\institute{Fondazione Bruno Kessler, Trento, Italy}

\titlespacing{\subsection}{0pt}{*0.9}{*0.9}

\begin{document}
\pagestyle{plain}
\maketitle

\begin{abstract}
This paper describes the \xsap safety analysis platform. \xsap
provides several model-based safety analysis features for
finite- and infinite-state synchronous transition systems. In
particular, it supports library-based definition of fault modes, an
automatic model extension facility, generation of safety analysis
artifacts such as Dynamic Fault Trees (DFTs) and Failure Mode and
Effects Analysis (FMEA) tables. Moreover, it supports probabilistic
evaluation of Fault Trees, failure propagation analysis using Timed
Failure Propagation Graphs (TFPGs), and Common Cause Analysis (CCA).
\xsap has been used in several industrial projects as verification
back-end, and is currently being evaluated in a joint R\&D Project
involving FBK and The Boeing Company.
\end{abstract}

 \input{files/introduction.tex}
\input{files/functionalities.tex}
\input{files/architecture.tex}
\input{files/applications.tex}

\input{files/conclusions.tex}

\spacing{0.95}
\bibliographystyle{splncs}
\bibliography{biblio}
\end{document}

%% file: files/introduction.tex
\section{Introduction}
In recent years, there has been a growing industrial interest in
model-based safety assessment techniques
(MBSA)~\cite{bozzano:03,bozzano:04,Joshi05:Dasc,SafetyAssessmentBook,DBLP:journals/cj/BozzanoCKNNR11}
and their application. These methods are based on a single
safety model of a system, and analyses are carried out with a high
degree of automation, thus reducing the most tedious and error-prone
activities that today are carried out manually.  Formal verification
tools based on model checking have been extended to automate the
generation of artifacts such as Fault Trees and FMEA tables, which are
required for certification of safety critical systems -- see,
e.g.,\cite{ARP4754A,ARP4761,ECSS}.

\xsap is a platform for Model-Based Safety Analysis (MBSA), which
provides a variety of features. First, it enables the definition of
fault modes, based on a customizable fault library. Second, it
implements automatic model extension, namely the possibility to
automatically extend a system model with the fault definitions
retrieved from the library. Third, it implements a full range of
safety analyses, including Fault Tree Analysis (FTA), Failure Mode and
Effects Analysis (FMEA), failure propagation analysis using Timed
Failure Propagation Graphs (TFPGs), and Common Cause Analysis (CCA).
Finally, \xsap implements a family of effective routines for such
analyses, based on state-of-the-art model checking techniques,
including BDD-, SAT- and \smt-based techniques.

\xsap is currently the core verification engine for many other tools,
including industrial ones. It has been used in several industrial
projects, including COMPASS~\cite{COMPASS-ITT},
AUTOGEF~\cite{AUTOGEF-ITT}, FAME~\cite{FAME-ITT} and
HASDEL~\cite{HASDEL-ITT}, funded by the European Space
Agency. Moreover, \xsap is currently being used in a joint research and
development project between FBK and The Boeing
Company~\cite{CAV-Boeing-paper}.

\xsap is being developed by FBK, and it is currently distributed with
a free license for academic research purposes and non-commercial
applications. \xsap can be downloaded from
\myurl{http://xsap.fbk.eu}.

\paragraph*{Related Work.}
The system closest to \xsap is the \fsap
platform~\cite{DBLP:journals/sttt/BozzanoV07}, which is no longer
maintained.  \xsap extends \fsap along several directions. First,
\fsap was limited to finite-state systems, while \xsap can handle
infinite-state ones. Second, \xsap provides more general and
customizable libraries to define fault modes and their dynamics and
new features, e.g., models and algorithms for failure propagation
analysis.  Third, \xsap implements a family of advanced routines for
safety analysis that extend those of \fsap in many respects -- namely,
the BDD-based Fault Tree generation routines described in
~\cite{ATVA07} are complemented by (different variants of) SAT-based
and SMT-based routines, and routines based on
IC3~\cite{DBLP:journals/corr/CimattiGMT13,CAV-algos-paper}. Finally,
while \fsap provides a graphical user interface, \xsap relies on an
interaction shell similar to \nuxmv, and models are expressed in
textual format, thus increasing the flexibility and possibility of
integration within other tools.

Some of the safety assessment functions of \xsap are used as a back-end
for the COMPASS
tool~\cite{DBLP:journals/cj/BozzanoCKNNR11,COMPASS-RESS} and its
extensions~\cite{AUTOGEF-DASIA,IMBSA-regular}.
There are two key differences with respect to the COMPASS
tools. First, \xsap provides a wider range of routines for Fault Tree
generation; second, \xsap implements a general model extension
mechanism, based on a library defining fault modes and their dynamics,
while in COMPASS the fault models must be modeled manually and
explicitly within the SLIM language.

Other platforms for MBSA are based on the Altarica language and
OCAS~\cite{ONERA1,Altarica3,SCP15}, on
Scade~\cite{DA04,Joshi05:SafeComp}, and on
Statemate~\cite{PBBSBH04,DBLP:conf/safecomp/PeikenkampCVBPH06}.  None
of them is publicly available.

Finally, in~\cite{Gudemann10} the
authors present a framework for model based safety analysis, that
provides transformations into different model checkers, including
NuSMV, but does not provide specialized algorithms.

\paragraph*{Structure of the paper.}  In
Sect.~\ref{sec:functionalities} and~\ref{sec:architecture} we describe
the functionalities and the architecture of \xsap.  In
Sect.~\ref{sec:applications} we briefly discuss its most successful
applications. In Sect.~\ref{sec:conclusions} we draw conclusions and
outline future directions.

%% file: files/functionalities.tex
\section{Functionality}
\label{sec:functionalities}
\begin{figure}[t]
\begin{center}
\includegraphics[width=1.0\columnwidth]{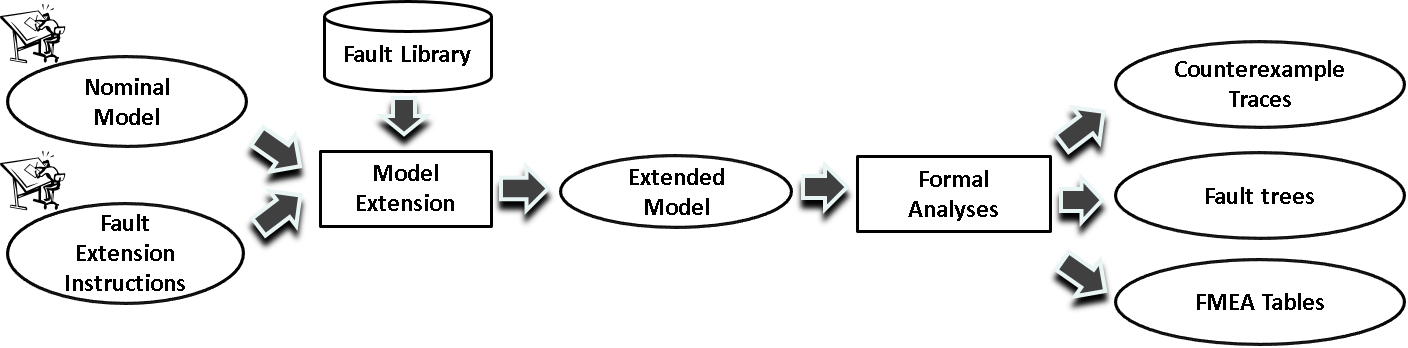}
\end{center}
\caption{The \xsap main flow.}
\label{fig:flow}
\end{figure}

In this section we describe the main features of \xsap.
Fig.~\ref{fig:flow} illustrates the main flow.

\subsection{Model Extension}
\label{sec:modelextension}
Model
extension~\cite{DBLP:journals/sttt/BozzanoV07,SafetyAssessmentBook} is
an automated process that, based on a specification of the possible
faults, returns a model (called {\em extended model}) that takes into
account faulty behaviors.  The model extension routine takes as input
the {\em nominal model} (describing behavior in absence of faults),
the {\em fault library} (containing templates for faults and their
dynamics) and the {\em fault extension instructions} (specifying
directives to instantiate the fault templates). Formal analyses can be
run on the extended model, in order to assess system behavior in
presence of faults.

The fault library of \xsap contains a comprehensive set of predefined
fault modes, including, e.g., different variants of {\em stuck at},
{\em random}, {\em conditional}, {\em ramp down}, and can be further
customized for any specific need. Moreover, a {\em local} and {\em
  global} dynamics libraries enable the definition of the dynamics of
faults (e.g., {\em permanent} or {\em sporadic}). The fault library
has been validated and extended to match the need of a significant
case study of industrial size~\cite{CAV-Boeing-paper}.
\begin{figure}[t]
\begin{center}
\begin{tabular}{ll}
{
\includegraphics[width=0.60\columnwidth]{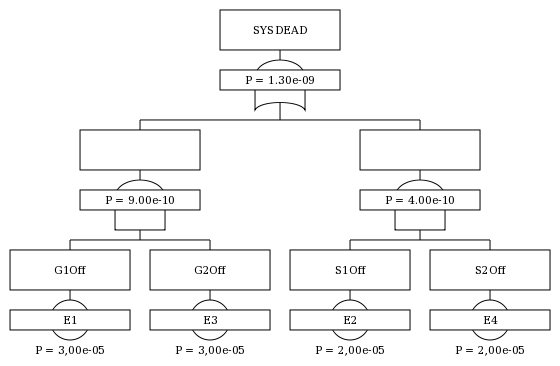}
}
&
{
\includegraphics[width=0.35\columnwidth]{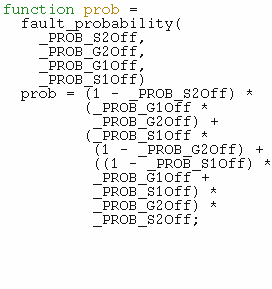}
}
\end{tabular}
\end{center}
\caption{An example FT and the associated symbolic probability.}
\label{fig:FT}
\end{figure}

\subsection{Safety Analysis}
\label{sec:safety}
\xsap supports the automatic generation of artifacts that are
typical of safety analysis, in particular Fault Trees and FMEA
tables~\cite{FTH2,ARP4754A,ARP4761}. A Fault Tree (FT) is a graphical
representation of the sets of possible causes of a given (undesired)
event (the root of the tree -- called {\em Top Level Event, TLE}). The
TLE is linked by means of logical gates (AND, OR) to the basic events
(faults). The minimal combinations of faults explaining the TLE are
called {\em Minimal Cut Sets (MCSs)}. Fig.~\ref{fig:FT} (left) shows a
sample FT for a battery sensor model: the FT links the TLE (a system
failure) with basic events (generator and sensor faults).  Finally,
\xsap can generate Dynamic Fault Trees
(DFTs)~\cite{MDCS98,DBLP:journals/tdsc/BoudaliCS10}, where a {\em
  priority AND} gate is used to identify order of precedence of
different events~\cite{bozzano:02}.

FMEA tables are a tabular representation of the causality
relationships between (sets of) faults and a list of properties
(representing undesired events, as in the case of FTs). \xsap also
supports the generation of Dynamic FMEA tables, where order of events
may be imposed.

\subsection{Common Cause Analysis}
\label{sec:cca}
Common Cause Analysis (CCA) is a necessary step of safety
assessment, that is often required by safety standards, see
e.g.,~\cite{ARP4754A,ARP4761}. It consists in evaluating the
consequences of events that may break the hypothesis of independence
of different faults. CCA aims at investigating possible dependencies,
and evaluates the consequences in terms of system
safety/reliability. \xsap enables the definition of events named {\em
  common causes}, which may trigger the occurrence of a set of
(dependent) faults. Such faults may follow a user-specified pattern,
e.g., {\em simultaneous} or {\em cascading} (subject to given temporal
constraints). For instance, debris caused by an engine burst (the
common cause) may cause multiple components of an aircraft to fail
simultaneously.  \xsap enables the evaluation of system reliability in
presence of common causes and the generation of FTs including them.

\subsection{Probabilistic Evaluation}
\label{sec:probability}
\xsap supports the probabilistic evaluation of Fault Trees. Given
numerical probabilities for the basic events and for the common
causes, \xsap computes probabilities for the intermediate nodes and
the TLE of a Fault Tree. With the exception of the
constituent faults of common causes, all faults are assumed to be
independent.

Furthermore, \xsap supports the symbolic computation of the formula
representing the probability of the TLE, given the probability of the
basic events. From such formula, \xsap produces the code in Python and
Matlab/Octave. This can be used to sample the reliability of a system
for different values of the fault probabilities, and plotted using
visualization tools (e.g., Matlab~\cite{ICECCS13}).  Fig.~\ref{fig:FT}
shows the FT with associated numerical probabilities, and the
symbolic formula representing the probability of the TLE.

\subsection{Failure Propagation Analysis}
\label{sec:tfpg}
\xsap supports analysis of failure propagation using Timed Failure
Propagation Graphs (TFPGs). A TFPG~\cite{Karsai2003} is a graph-like
model that accounts for the temporal progression of failures in
dynamic systems and for the causality between failure effects, taking
into consideration time delays, system reconfiguration and sensor
failures. TFPGs may be seen as a more fine-grained model w.r.t. DFTs, and
can be used to support important run-time activities such as diagnosis
and prognosis~\cite{Hayden2006,Ofsthun2007,IMBSA-regular}.

\begin{figure}[t]
\begin{center}
\resizebox{0.95\textwidth}{!}{\input{figs/tfpg_example}}
\end{center}
\caption{An example TFPG.}
\label{fig:TFPG}
\end{figure}
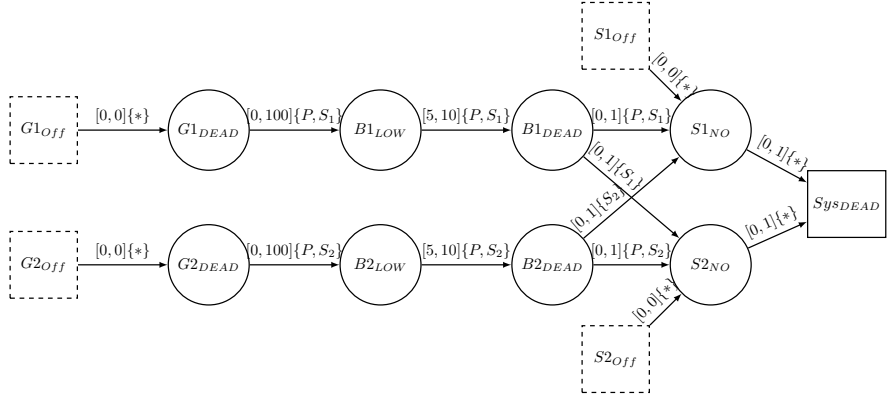

The nodes of a TFPG represent either {\em failures} or {\em
  discrepancies} (representing anomalous behaviors). Edges represent
propagation links; they are labeled with timing information (bounding
the minimum and maximum propagation time) and modes (information on
system modes enabling the propagation, e.g., operational modes of a
spacecraft). Discrepancies may be given either AND or OR semantics --
in the former case all incoming edges must be active in order for the
failure to propagate, in the latter case any of them suffices.
Fig.~\ref{fig:TFPG} shows a sample TFPG for the battery sensor
model. Nodes without incoming edges (shown with dotted lines) are
failures, whereas the remaining nodes are discrepancies (AND
discrepancies are drawn as boxes, and OR discrepancies as
circles). Edges are labeled with propagation bounds (in square
brackets) and system modes (in curly brackets).

\xsap supports modeling and validation of TFPGs~\cite{TACAS14}. In
particular, behavioral validation has the purpose to check whether a
TFPG is a complete abstraction of a given system model (i.e., it
contains at least as many behaviors as the system it represents).
Finally, \xsap supports automatic synthesis of a TFPG from a model,
given a set of failures and discrepancies (currently, only the
structure of the graph is synthesized). The integration of the TFPG
validation features described in~\cite{AAAI15} is under way.

%% file: figs/tfpg_example.tex
\begin{tikzpicture}
  [
    ornode/.style={circle,draw=black,thick,minimum size=1.5cm, text width=1.5cm, align=center},
    andnode/.style={rectangle,draw=black,thick,minimum size=1.5cm},
    fault/.style={rectangle,dashed,draw=black,thick,minimum size=1.5cm},
    monitor/.style={diamond,draw=black,thick, minimum size=1.5cm},
    conn/.style={thick,solid,-latex},
    monitorconn/.style={thick,dashed,-latex},
    every node/.style={node distance=2cm}
  ]

  \node[fault] (gen1off) {$G1_{Off}$};
  \node[fault] (gen2off) [below=1.5cm of gen1off] {$G2_{Off}$};

  \node[ornode] (g1dead) [right=of gen1off] {$G1_{DEAD}$};
  \node[ornode] (g2dead) [right=of gen2off] {$G2_{DEAD}$};

  \node[ornode] (b1low) [right=of g1dead] {$B1_{LOW}$};
  \node[ornode] (b2low) [right=of g2dead] {$B2_{LOW}$};

  \node[ornode] (b1dead) [right=of b1low] {$B1_{DEAD}$};
  \node[ornode] (b2dead) [right=of b2low] {$B2_{DEAD}$};

  \node[ornode] (s1wo) [right=1.7cm of b1dead] {$S1_{NO}$};
  \node[ornode] (s2wo) [right=1.7cm of b2dead] {$S2_{NO}$};

  \node[fault] (sens1off) [above left=1cm of s1wo] {$S1_{Off}$};
  \node[fault] (sens2off) [below left=1cm of s2wo] {$S2_{Off}$};

  \node[andnode] (systemdead) [below right=0.25cm and 1.5cm of s1wo] {$Sys_{DEAD}$};


  \path (gen1off) edge [conn] node [anchor=south,sloped] {$[0,0] \{*\}$} (g1dead);
  \path (g1dead) edge [conn] node [anchor=south,sloped] {$[0,100] \{P,S_1\}$} (b1low);
  \path (b1low) edge [conn] node [anchor=south,sloped] {$[5,10] \{P,S_1\}$} (b1dead);
  \path (b1dead) edge [conn] node [anchor=south,sloped] {$[0,1] \{P,S_1\}$} (s1wo);
  \path (sens1off) edge [conn] node [anchor=south,sloped] {$[0,0] \{*\}$} (s1wo);
  \path (s1wo) edge [conn] node [anchor=south,sloped] {$[0,1] \{*\}$} (systemdead);

  \path (gen2off) edge [conn] node [anchor=south,sloped] {$[0,0] \{*\}$} (g2dead);
  \path (g2dead) edge [conn] node [anchor=south,sloped] {$[0,100] \{P,S_2\}$} (b2low);
  \path (b2low) edge [conn] node [anchor=south,sloped] {$[5,10] \{P,S_2\}$} (b2dead);
  \path (b2dead) edge [conn] node [anchor=south,sloped] {$[0,1] \{P,S_2\}$} (s2wo);
  \path (sens2off) edge [conn] node [anchor=south,sloped] {$[0,0] \{*\}$} (s2wo);
  \path (s2wo) edge [conn] node [anchor=south,sloped] {$[0,1] \{*\}$} (systemdead);

  \path (b1dead) edge [conn] node [anchor=south,sloped,near start] {$[0,1] \{S_1\}$} (s2wo);
  \path (b2dead) edge [conn] node [anchor=south,sloped,near start] {$[0,1] \{S_2\}$} (s1wo);


\end{tikzpicture}

%% file: files/architecture.tex
\section{Architecture and Implementation}
\label{sec:architecture}
The architecture of \xsap is built around the \nuxmv symbolic model
checker~\cite{cav2014-nuxmv}, from which it inherits all the
functionalities.
\nuxmv is an extension of \nusmv, and supports the verification of
finite- and infinite-state systems, by means of advanced SAT- and
SMT-based model checking techniques.
\nuxmv provides to \xsap the basic infrastructure, e.g., the symbol
table, the flattening of the design, the Boolean encoding of scalar
variables, the representation of state machines and temporal formulae,
and the basic model checking algorithms.

On top of this, \xsap features the following blocks.
\emph{Model Extension} includes the library of fault modes, a parser
for the fault extension instruction language, and the model extension.
\emph{Minimal cut sets computation} is realized by way of routines for
parameterized model checking, using the model checking primitives of
\nuxmv as building blocks.
\emph{Fault Trees} can be generated/stored/retrieved either in XML or
in a standard textual (tab-separated) format supported by commercial
tools, such as FaultTree+~\cite{FTPlus}.
The management of \emph{FMEA tables} is isolated in a separate module.
Support for \emph{Time Failure Propagation Graphs} is based on XML and
textual formats. The textual format has been conceived to enable
editing in a human-readable form -- \xsap provides conversion from
textual to XML and vice versa.
\emph{Syntax Directed Editors} (SDEs) are available for editing models,
fault extension instructions, and TFPGs.
Finally, the \emph{Visualization} module contains graphical viewers
that can be used to display the safety analysis artifacts: an FT
Viewer and a TFPG viewer are available for the analysis of FTs and
TFPGs, respectively.

\xsap has been developed in C and in C++ for the internal modules,
while Python is used for model extension and TFPG manipulation. The
viewers are based on the PyGTK, Goocanvas, PyGraphviz and Matplotlib
libraries.
\xsap compiles and executes on the most widely used Operating Systems
(OSs) and architectures, namely: Linux, MS Windows, and MacOS
X. Porting to other OSs is also possible (although not tested yet).

%% file: files/applications.tex
\section{Applications}
\label{sec:applications}

The \xsap platform has been used in a wide range of applications, both
at academic and at industrial level, and in several domains,
including avionics and aerospace, railway and industrial control.
\xsap is the successor of \fsap~\cite{DBLP:journals/sttt/BozzanoV07}
and retains all its features. \fsap has been developed within the
ESACS~\cite{ESACS}, ISAAC~\cite{ISAAC}, and MISSA~\cite{MISSA}
projects. It pioneered the ideas of model extension and model-based
safety assessment, and was applied for safety assessment of avionics
system~\cite{bozzano:03,bozzano:04,bozzano:06}.

\xsap has been widely used in several industrial projects with the
European Space Agency (ESA). It is the back-end of the COMPASS
tool~\cite{DBLP:conf/cav/BozzanoCKNNRW10,DBLP:journals/cj/BozzanoCKNNR11,COMPASS-RESS},
developed within the COMPASS~\cite{COMPASS-ITT},
AUTOGEF~\cite{AUTOGEF-ITT,AUTOGEF-DASIA},
FAME~\cite{FAME-ITT,IMBSA-regular} and HASDEL~\cite{HASDEL-ITT}
projects funded by ESA.

Currently, \xsap is being used by Boeing~\cite{CAV-Boeing-paper}. The
Boeing Company has evaluated \xsap in the context of a joint research
and development project in the areas of model-based safety assessment,
verification and validation, and contract-based design. The purpose of
this project is to demonstrate the usefulness and suitability of
model-based safety assessment techniques for improving the overall
process in terms of robustness and cost-effectiveness, and for
certification purposes. In this context, \xsap has been used to model
an industrial-size case study~\cite{CAV-Boeing-paper,AIR6110} and
thoroughly evaluated in an industrial setting.

%% file: files/conclusions.tex
\section{Conclusions and Future Work}
\label{sec:conclusions}
In this paper we presented \xsap, a state-of-the-art platform for
model-based safety analysis, providing a full range of
functionalities, based on symbolic model checking techniques.  We
described the architecture of \xsap and its industrial applications.

The symbolic technologies implemented in \xsap provide significant
advances also in terms of scalability. We refer to~\cite{SCP15} for a
comparison with Altarica/OCAS (carried out using a license courtesy of
Dassault Aviation), and to~\cite{CAV-algos-paper} for an exhaustive
evaluation of the novel routines implemented in \xsap.

As future work, we intend to extend \xsap in several
directions. First, we want to incorporate Contract-Based Safety
Assessment (CBSA) techniques~\cite{ATVA14}, enabling the generation of
hierarchical FTs following the design structure. Moreover, we wish to
incorporate the routines for evaluation of reliability architectures
we developed in~\cite{HVC13}. Finally, a significant extension will
concern the definition of observability information in the model and
the addition of related functionalities, such as diagnosability
analysis and Fault Detection, Fault Isolation and Fault Recovery
(FDIR) analysis~\cite{TACAS14}.